\documentclass{article} \textheight= 240mm \textwidth=160mm \topmargin = -1 cm
\usepackage{makeidx}
\usepackage{graphicx}
\usepackage{amssymb}
\input epsf
\begin{document}

\title{\textbf{Thermodynamic theory of kinetic overshoots}}

\author{Alexander N. Gorban\thanks{ag153@le.ac.uk}\\
Centre for Mathematical Modelling, University of Leicester, UK,
\\ {\it and} Institute of Computational Modeling RAS \\ Krasnoyarsk Russia \\
Gregory S. Yablonsky\thanks{gy@che.wustl.edu} \\
Department of Chemical Engineering, Washington University \\
St. Louis, USA }

\date{}

\maketitle

\begin{abstract}
Everything that is not prohibited is permissible. So, what is
prohibited in the course of chemical reactions, heat transfer and
other dissipative processes? Is it possible to ``overshoot" the
equilibrium, and if yes, then how far? Thermodynamically allowed and
prohibited trajectories of processes are discussed by the example of
effects of equilibrium encircling. The complete theory of
thermodynamically accessible states is presented. The space of all
thermodynamically admissible paths is presented by projection on the
``thermodynamic tree", that is the tree of the related thermodynamic
potential (entropy, free energy, free enthalpy) in the balance
polyhedron. The stationary states and limit points for open systems
are localized too.
\end{abstract}

\section{Introduction}

In the beginning was $A$, the chemical reagent $A$, $100 \%$ of $A$
in a closed vessel. Then $A$ began to transform in $B$. The
substance $B$ transforms in $A$, as well as $A$ transforms in $B$.
The conditions are stationary; the reactions rates are proportional
to concentrations. The system goes to equilibrium. Let the
concentration of $A$ and $B$ be equal at this equilibrium (simply
for definiteness). One could ask a question: Is it possible that the
amount of $B$ exceed $50 \%$ during the way to equilibrium? The
answer is obvious: it is impossible. When concentration of $B$
reaches $50 \%$, the motion stops, because it is equilibrium state.
The system is one-dimensional, and one coordinate (for example, the
concentration of $A$) describes the state of the system in full (for
fixed external conditions, for example for fixed volume and
temperature). On the line it is impossible to walk round the
equilibrium.

The answer changes, if in the system a third reagent $C$ is present:
The equilibrium encircling is possible if the dimension is more than
$1$. The second question arises immediately: how far is it possible
to walk around the equilibrium? Is it possible to reach $100 \%$ or
$90 \%$ of $B$? How far the system can go along the
thermodynamically admissible paths, that are the continuous curves,
which satisfy the balances (conservation lows) and the second law
(the entropy of isolated system should grow monotonically, i.e. the
related thermodynamic potential of our system should change
monotonically, in proper direction). Let the equilibrium
concentrations of $A$, $B$, and $C$ be equal, all these reagents can
transform to each other, and the system is thermodynamically
perfect. In this case, on thermodynamically admissible paths the
concentration of $B$ cannot exceed the upper boundary
$b_{\max}\approx 77.3 \%$  on the way from the state $A=100 \%$ to
equilibrium \cite{obh}. This {\it global} thermodynamic estimation
follows from the {\it local} condition (the balances and the second
law) and from the continuity of thermodynamically admissible paths.

The questions about global thermodynamical restrictions for kinetic
behavior arose several times \cite{obh,chmms,varna,IChE,Shi,Shi2}.
The interest to this problem revives again and again (see, for
example, \cite{Vud,Kag}) because thermodynamic data are much more
reliable than kinetic constants, ant it is important to extract all
the possible information about dynamic behavior from thermodynamics,
both from the practical (industrial) and the scientific points of
view.

The goal of our talk is to analyze the main theoretical construction
that allow to solve the problem of global thermodynamical
restrictions on the system dynamics in an explicit form. It is the
problem of integration  of special differential inclusions: the time
derivatives comply with the thermodynamic inequalities and balance
equalities at each state, and the whole motion satisfy some global
restrictions. It appears that the thermodynamic conditions allow the
explicit integration.

\section{Thermodynamic tree}

Let us consider the chemical system dynamics in the composition
space. Coordinates in this space are amounts of reagents. For
systems with constant volume we can use the concentration space. The
linear conservation laws together with the positivity conditions
define a convex polyhedron $D$ in the composition space. For
chemical reaction under given condition there exists a thermodynamic
Lyapunov function. It is the appropriate thermodynamic potential.
For example, for $V,T={\rm const}$ it is the Helmholtz free energy
$F$, and for $P,T={\rm const}$ it is the Gibbs free energy (the free
enthalpy) $G$. We use the notation $G$ for any thermodynamic
Lyapunov function. The function $G$ is assumed to be {\it continuous
and strictly convex} in $B$. This assumption is crucial, because for
non-convex function $G$ computation difficulties of thermodynamic
analysis increase drastically. The global minimum of $G$ in $D$
belongs to interior $G$. It is the equilibrium point.

{\it Thermodynamically admissible path}  is such a continuous
function $\varphi : [0,1]\rightarrow D$ that the function $G(\varphi
(x))$ monotonically decrease (non-increase) on $[0,1]$. The state $y
\in D$ is {\it thermodynamically accessible} from the state $x \in
D$ ($x\succ y$), if there exists such a thermodynamically admissible
path $\varphi$ that $\varphi(0)=x$ and $\varphi(1)=y$. The states
$x,y \in D$ are {\it thermodynamically equivalent} ($x \sim y$), if
$x\succ y$ and $y\succ x$. In order to study the structure the
thermodynamic accessibility let us glue the thermodynamically
equivalent states: {\bf Thermodynamic tree is the factor-space}
$D/\sim$. This space is a one-dimensional continuum with finite
number of branching points, that is the tree. The function $G$ is
constant on the classes of thermodynamic equivalence, hence, we can
define $G$ on $D/\sim$. Similarly, the thermodynamic order $\succ$
can be defined on $D/\sim$, and  $y$ is thermodynamically accessible
from $x$ if and only if for their images on the thermodynamic tree
$(x/\sim) \succ (y/\sim)$. The set of all states thermodynamically
accessible from the given $x \in D$ is the preimage in $D$ of the
monotone  path going on the thermodynamic tree from the point
$x/\sim$ to the equilibrium.

\section{How to construct the thermodynamic tree?}

In this section we consider three main computational problems:
\begin{enumerate}
\item{How to construct the thermodynamic tree for given $D$ and $G$.}
\item{How to find the image of the given state $x\in D$ on the thermodynamic
tree.}
\item{How to find maximum of linear function on a class of thermodynamically
equivalent states, that is, on the preimage in $D$ of the point from
the thermodynamic tree, and on the preimage of a monotone path.}
\end{enumerate}

The solution of the third problem gives us the thermodynamically
admissible extremal values of various characteristics on the way to
equilibrium. After solution of the first two problems this problem
turns to the standard problem of optimization: to find a maximum of
a linear function on a convex set.

Let $D_1$ be the one-skeleton of $D$, that is the union of all edges
of $D$ (including the vertexes). For each $a$ ($ \min_{x\in D} G(x)
< a < \max_{x\in D} G(x)$) there is one-to-one correspondence
between connected components of sets $\{x\in D_1 | G(x) > a \}$ and
$\{z \in D /\sim | G(z) > a \}$. A connected component of $\{x\in
D_1 | G(x) > a \}$ maps by the natural projection $D \rightarrow  D
/\sim$ onto correspondent connected component of $\{z \in D /\sim |
G(z) > a \}$. Hence, for convex functions $G$ the solution of  the
first problem depends only on two sets of numbers: the values of $G$
in vertexes, and minimal values of $G$ on edges of $D$.

For solution of the second problem, it is sufficient to find such a
vertex $v$ that $v \succ x$. In this case the correspondent
thermodynamically admissible path can be chosen as a segment of a
straight line. For each vertex $v$ and the number $a$ ($ \min_{x\in
D} G(x) < a < G(v)$) the following two conditions, $v \succ x$ and
$G(x)=a$, define the point on the thermodynamic tree uniquely.

\section{Localization of stationary states}

Thermodynamics allows us to localize the domain in the concentration
space where the stationary states for open system can appear. This
estimation is rather simple: in this region the external flow should
produce $G$, whereas the intrinsic dynamics decrease $G$. Let the
kinetic equation have the form
$$\dot{N}= Vw(c) + v_{\rm in}c_{\rm in} - v_{\rm out}c,$$
where $N$ is composition vector, $c$ is vector of concentrations,
$V$ is the system volume, $w$ describes all the intrinsic processes,
$v_{\rm in}$ and $v_{\rm out}$ are input and output velocities, and
$c_{\rm in}$ is input vector of concentrations. The point $c$ might
be a stationary point of the open system only in the case when the
following inequality is true:
$$\left(\frac{v_{\rm in}}{v_{\rm out}}c_{\rm in} - c, \nabla
G(c)\right) \geq 0 ,$$ where (\, ,\, ) is the standard scalar
product.

Is it possible to localize the possible general limit points too
\cite{obh}. The available information about stoichiometric reaction
mechanism can be used for refinement of  these estimations.

\section{Discussion}

As it is well known, the traditional thermodynamic analysis gives
many possibilities to analyze the complex chemical process without
knowing a kinetic model:
\begin{itemize}
\item{to find the allowed direction of the chemical process;}
\item{to calculate the equilibrium composition of the complex
reaction mixture;}
\item{to apply relationships between parameters of the complex
chemical reaction.}
\end{itemize}
The strong advantage of the traditional approach is its simplicity
and reliability of data, which it is based on. At the same time, it
has the obvious limitation. Using this approach it is impossible to
estimate the dynamic of the complex chemical process ``beyond the
equilibrium".

In difference from the traditional approach, the advanced
thermodynamical analysis of kinetics (let us call it ``TAK") allows
to make the further significant step in analysis with no kinetic
model. Using TAK, it became possible to estimate an efficiency
(selectivity, yield) for the complex chemical reaction ``far from
equilibrium", i.e. under non-steady-state conditions in the closed
system or under steady-state conditions (or in more complicated
attractor regime) in the open system.

Many toy and not only toy example are presented now in
\cite{obh,chmms,Shi2,Kag}. Using TAK, the following problems can be
solved:
\begin{enumerate}
\item{Knowing the initial composition of the complex mixture,
to estimate the characteristics of the process ``far from
equilibrium" and answer the question, is it possible to achieve the
desired values of process characteristics.}
\item{Assuming the desired characteristics of the process,
to estimate the corresponding domain of the initial composition.}
\end{enumerate}

\end{document}